\documentclass{svjour3mod} 

\usepackage{graphicx}
\usepackage{mathptmx}
\usepackage{amsmath,amssymb,amsfonts}

\newcommand{\Order}{\mathcal{O}}

\journalname{Hyperfine Interact.}

\begin{document}

\title{Charge-symmetry-breaking nucleon form factors}

\author{Bastian Kubis}

\institute{Bastian Kubis \at
              Helmholtz-Institut f\"ur Strahlen- und Kernphysik (Theorie) and  
              Bethe Center for Theoretical Physics, \\
              Universit\"at Bonn, 
              D-53115 Bonn, Germany\\
              \email{kubis@hiskp.uni-bonn.de}
}

\date{}

\maketitle

\begin{abstract}
A quantitative understanding of charge-symmetry breaking is an increasingly important ingredient 
for the extraction of the nucleon's strange vector form factors.  
We review the theoretical understanding of the charge-symmetry-breaking form factors, 
both for single nucleons and for ${{}^4{\rm He}}$.

\end{abstract}

\section{Introduction}

The investigation of strangeness contributions to static properties
of the nucleon is particularly interesting as it gives unambiguous access to 
low-energy manifestations of virtual or sea quark effects.
Different strangeness currents of the form $\bar s \Gamma s$
test the strangeness component of different nucleon observables,
such as mass ($\Gamma = 1$), spin ($\Gamma = \gamma_\mu\gamma_5$), or 
magnetic moment ($\Gamma = \gamma_\mu$).  
Here we are concerned with the magnetic moment only, or, more generally,
with the nucleon form factors of the vector current.

\begin{sloppypar}
The Standard Model provides access to two different flavor combinations of the
three light quark contributions to the 
electric ($G_E$) and magnetic ($G_M$) form factors due to the electromagnetic
and the weak vector currents, 
\begin{eqnarray}
G_{E/M}^{\gamma,p} &=& \frac{2}{3}G_{E/M}^u - \frac{1}{3}\bigl(G_{E/M}^d  + G_{E/M}^s \bigr) ~,  \\
G_{E/M}^{Z,p} &=& \Bigl(1-\frac{8}{3}\sin^2\theta_W\Bigr) G_{E/M}^u 
             - \Bigl(1-\frac{4}{3}\sin^2\theta_W\Bigr) \bigl( G_{E/M}^d +G_{E/M}^s \bigr) ~. 
\end{eqnarray} \vskip -1mm \noindent
In order to obtain a full flavor decomposition of the vector current,
one therefore has to invoke isospin (or charge) symmetry in the form 
\begin{equation}
G_{E/M}^{u,n}=G_{E/M}^{d,p} ~, \quad G_{E/M}^{d,n}=G_{E/M}^{u,p} ~,
\end{equation}
and use the neutron electromagnetic form factors as the third input.
These relations are at the heart of the extensive experimental program
to extract the \emph{weak} form factors of the proton $G_{E/M}^{Z,p}$
from parity-violating electron scattering~\cite{Aniol:2000at,Ito:2003mr,Spayde:2003nr,Maas:2004ta,Maas:2004dh,Aniol:2005zg,Armstrong:2005hs,Acha:2006my,Baunack:2009gy,Androic:2009zu}.
If one relaxes this assumption and allows for charge-symmetry breaking, however,
the relation between weak vector form factors, electromagnetic form factors
of proton and neutron, and strangeness is complicated by an additional term, 
\begin{equation}
G_{E/M}^{Z,p} = \bigl(1-4\sin^2\theta_W \bigr) G_{E/M}^{\gamma,p} - G_{E/M}^{\gamma,n} - G_{E/M}^s
- G_{E/M}^{u,d} ~, 
\end{equation}  
where
$G_{E/M}^{u,d} = 2/3( G_{E/M}^{d,p} - G_{E/M}^{u,n} )
-1/3 ( G_{E/M}^{u,p} - G_{E/M}^{d,n})$\,.
In other words, the charge-symmetry-violating form factors $G_{E/M}^{u,d}$ generate
``pseudo-strangeness'', and in order to reliably extract strangeness effects,
the former have to be calculated from theory.
This is becoming a necessity in particular in view of the increasingly tighter bounds
on strangeness deduced from experiment~\cite{Aniol:2000at,Ito:2003mr,Spayde:2003nr,Maas:2004ta,Maas:2004dh,Aniol:2005zg,Armstrong:2005hs,Acha:2006my,Baunack:2009gy,Androic:2009zu}.
\end{sloppypar}

\section{Theory of charge-symmetry-breaking form factors}

The isospin-breaking form factors $G_{E/M}^{u,d}$ have been addressed in various models
of the strong interactions, in particular in constituent quark~\cite{Dmitrasinovic:1995jt,Miller:1997ya}
or light-cone meson--baryon models~\cite{Ma:1997gh} (see also Ref.~\cite{Lewis:2006rd} 
for a comparative review).  These are afflicted by several problems: first, as in general
with model calculations, it is extremely difficult to quantify the inherent
uncertainties; second, the symmetries of the Standard
Model may be violated.  E.g., quark models~\cite{Dmitrasinovic:1995jt,Miller:1997ya}
predict $G_{E/M}^{u,d}(t=0)=0$ for the 
charge-symmetry breaking magnetic moment, which is only due to a specific symmetry of the 
quark model wave function employed, not, as we shall see below, due to a symmetry of the Standard Model,
and may consequently lead to an underestimation of isospin-breaking effects in particular at small $t$.

Chiral perturbation theory (ChPT)~\cite{weinbergchpt,glannphys}, on the other hand,
is ideally suited for an analysis of isospin violation.  
It is tailor-made to analyze the dependence of low-energy observables
on quark masses, in particular on the light quark mass difference $m_u-m_d$,
and the consistent inclusion of electromagnetic effects is also 
well-understood~\cite{urech}.
As the isospin-violating form factors can be calculated in SU(2) ChPT,
they are not affected by convergence problems to the extent
the strangeness form factors are
(see Ref.~\cite{Kubis:2005cy} for a review on the latter).

Particular emphasis will be put on the analysis of the leading moments
of the isospin-violating form factors, magnetic moment as well as
electric and magnetic radius terms, 
\begin{equation}  
G_E^{u,d}(t) = \rho_E^{u,d} t + \Order(t^2) ~, \qquad
G_M^{u,d}(t) = \kappa^{u,d} + \rho_M^{u,d} t + \Order(t^2)  ~. 
\end{equation}
The two radius terms are unaffected by low-energy constants up to
leading ($\rho_E^{u,d}$) and next-to-leading ($\rho_M^{u,d}$)
order and can be expressed entirely in terms of the 
neutron-to-proton mass difference $\Delta m=m_n-m_p$~\cite{Lewis:1998iu},
with the result~\cite{Kubis:2006cy}  
\begin{equation} 
\rho_E^{u,d} = \frac{5\pi C}{6M_\pi m_N} ~, \quad
\rho_M^{u,d} = \frac{2C}{3M_\pi^2}\biggl\{1-\frac{7\pi}{4}\frac{M_\pi}{m_N}\biggr\} ~, \quad
C = \frac{g_A^2 m_N \Delta m}{16\pi^2F_\pi^2} ~. \label{eq:loop}
\end{equation}
Note that these expressions for the radii are entirely non-analytic in the quark masses.
Their simplicity is remarkable: 
the pion mass difference $M_{\pi^+}^2-M_{\pi^0}^2$ cannot play a role as it only 
breaks charge \emph{independence}, not charge symmetry; furthermore,
up to $\Order(p^4)$ for $G_E^{u,d}$ and 
$\Order(p^5)$ for $G_M^{u,d}$, no photon loops contribute, nor
are there any two-loop effects.  
In fact, ChPT is more predictive here than for the usual electromagnetic 
form factors of the nucleon (see e.g.\ Ref.~\cite{Kubis:2000zd}), 
precisely because polynomial counterterm contributions
must be suppressed by isospin-breaking factors $m_u-m_d$ or $e^2$, and hence
by two orders in the chiral expansion.

In order to complete the chiral representation, we have to estimate
the combination of low-energy constants entering $\kappa^{u,d}$. 
This is done by invoking the principle of resonance saturation:
low-energy constants in effective theories incorporate the effects of heavier states
not included in the theory as explicit degrees of freedom.
In our case, 
the relevant contributions are provided by vector mesons including
$\rho-\omega$ mixing~\cite{Kubis:2006cy}: 
\begin{eqnarray}
G_E^{u,d}(t) \Bigr|_{\rm mix} &=& 
   \frac{\Theta_{\rho\omega}\,t}{M_V(M_V^2-t)^2}
   \biggl[
   \Bigl(1 + \frac{\kappa_\omega M_V^2}{4m_N^2} \Bigr)g_\omega F_\rho
 - \Bigl(1 + \frac{\kappa_\rho M_V^2}{4m_N^2} \Bigr)g_\rho F_\omega 
   \biggr] ~, \nonumber \\
G_M^{u,d}(t) \Bigr|_{\rm mix} &=&
     \frac{\Theta_{\rho\omega}}{M_V(M_V^2-t)^2}
     \Bigl[
     \bigl(t + \kappa_\omega M_V^2\bigr)g_\omega F_\rho
   - \bigl(t + \kappa_\rho M_V^2\bigr)g_\rho F_\omega
     \Bigr] ~. \label{eq:mix}
\end{eqnarray} 
The necessary couplings can be extracted from experimental data
within certain errors.  Such an inclusion of phenomenological vector-meson
contributions has been shown to cure the main deficits of a chiral one-loop
representation of the usual nucleon electromagnetic form factors~\cite{Kubis:2000zd},
and is expected to work even better here due to the stronger suppression of even higher mass
states in the mixing amplitudes.
Still, the uncertainties in particular in the vector-meson--nucleon coupling constants
$g_{\rho/\omega}$, $\kappa_{\rho/\omega}$~\cite{Belushkin:2006qa} limit the precision of the
prediction for the isospin-violating form factors; see Ref.~\cite{Kubis:2006cy}
for a detailed discussion.

Although, strictly speaking, the chiral representation of the form factors 
to $\Order(p^4)$ ($G_E^{u,d}$) and $\Order(p^5)$ ($G_M^{u,d}$) only requires
an estimate for the low-energy constant entering the isospin-violating magnetic moment,
the representation Eq.~\eqref{eq:mix} also allows to assess higher-order counterterms
contributing to the radii.  Numerically one finds that, although formally subleading,
the large vector-meson couplings tend to overwhelm the loop contributions Eq.~\eqref{eq:loop} in the radii, 
which scale with the (small) nucleon mass difference.  We therefore include the full mixing amplitudes
in the final predictions.

\begin{figure*}
\begin{center}
\includegraphics[width=0.49\linewidth]{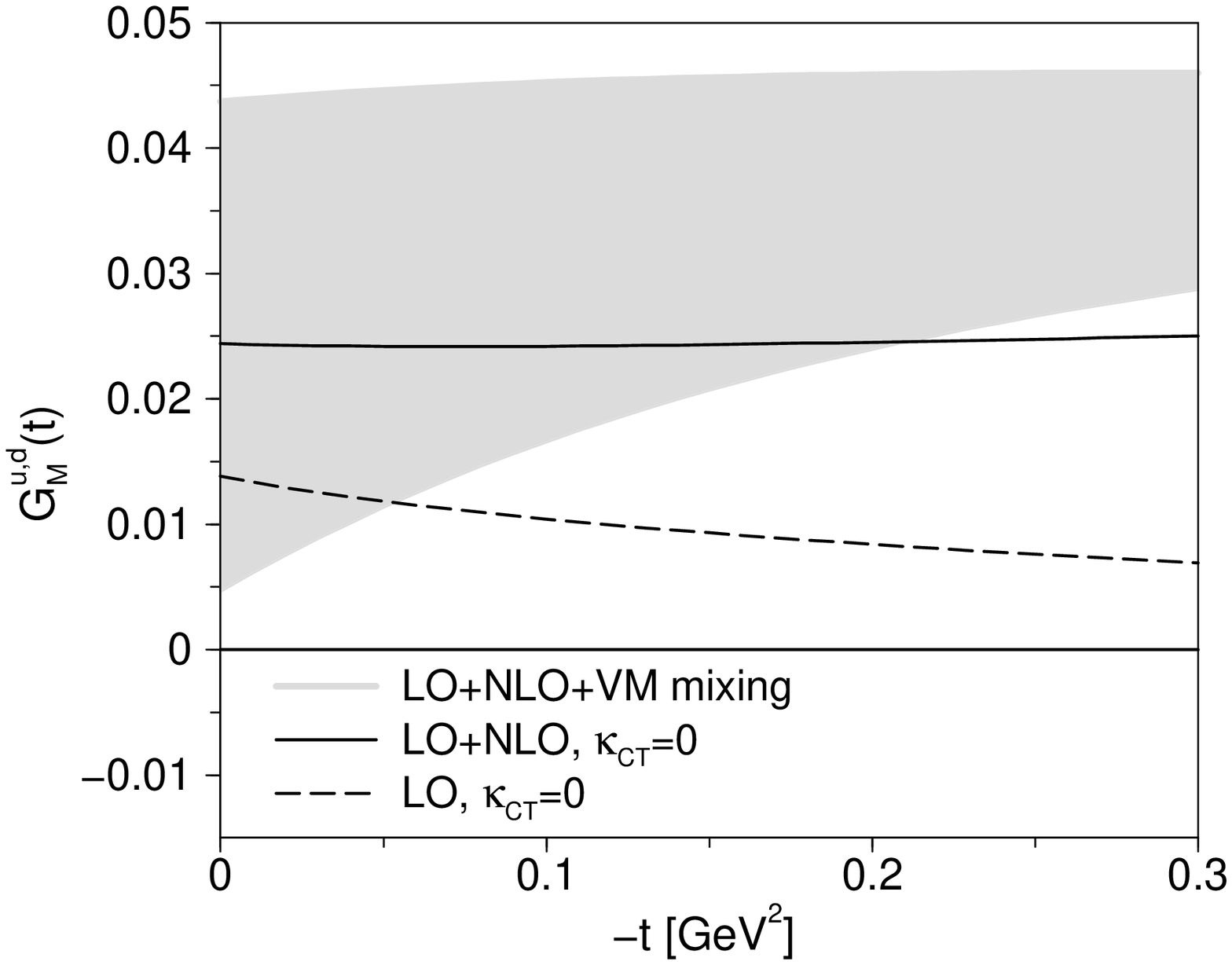} \hfill
\includegraphics[width=0.49\linewidth]{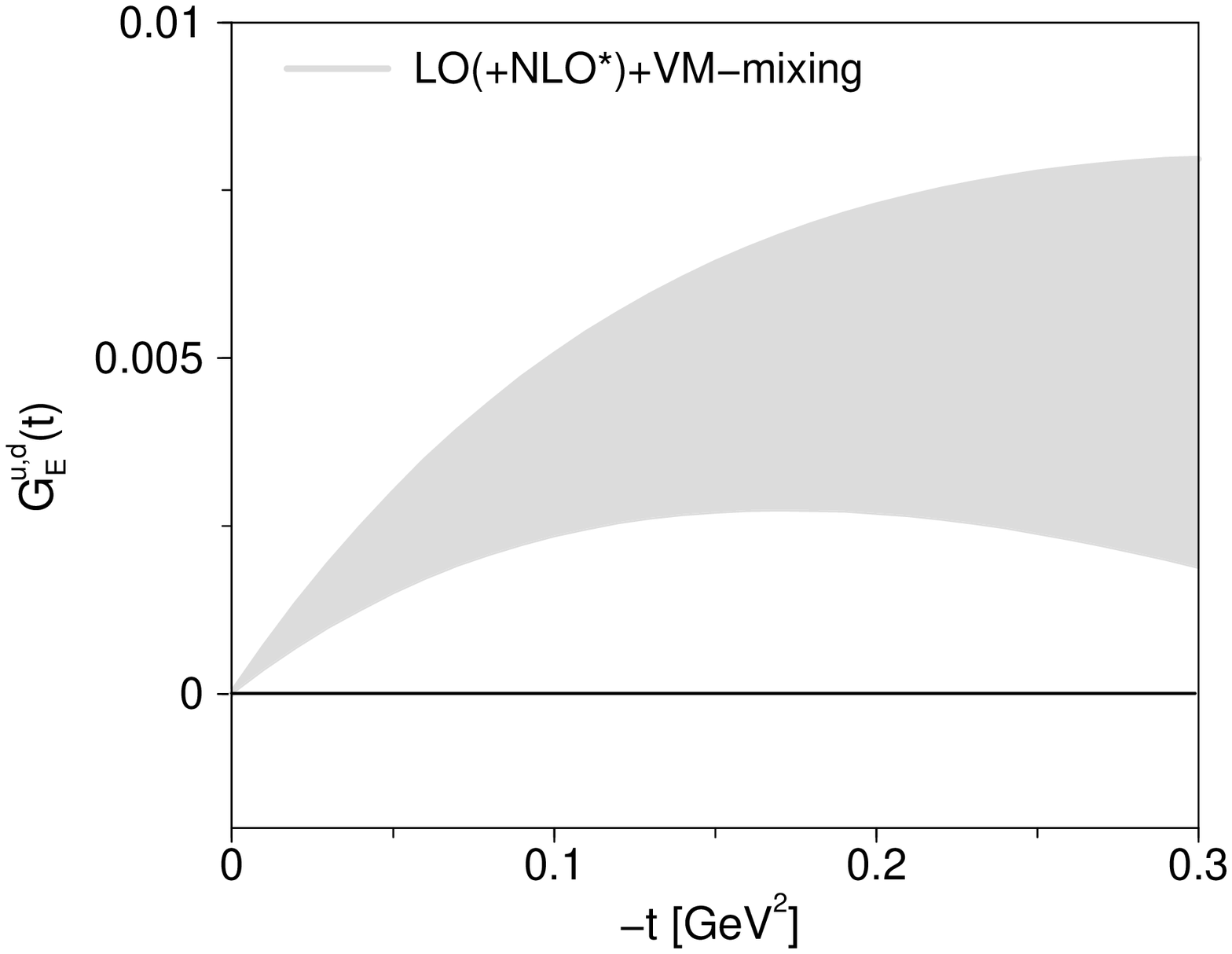} 
\caption{Charge-symmetry-breaking magnetic (left) and electric (right) form factors.
The grey bands denote the combined theoretical uncertainty due to various input parameters.
Figures taken from Ref.~\cite{Kubis:2006cy}.} \label{fig:ffresults}
\end{center} 
\end{figure*}
\begin{sloppypar}
The total results for the charge-symmetry-breaking form factors are shown in Fig.~\ref{fig:ffresults}.
The error bands combine an estimate of chiral corrections at higher order with the
above-mentioned uncertainties in the input coupling constants for the resonance contributions.
Although these combined uncertainties are sizeable, several conclusions can still be drawn:
the effects of isospin breaking remain at the percent level; the $t$-dependence 
of the form factors is rather moderate in the low-energy region.
We note that the symmetries of the Standard Model do \emph{not} dictate $\kappa^{u,d}$ to vanish,
indeed we find $G_M^{u,d}(0) \neq 0$.
\end{sloppypar}

The validity of these results, and in particular of the prediction for 
$\kappa^{u,d}=0.025\pm 0.020$~\cite{Kubis:2006cy}, has been criticized as an
``extreme estimate'' in Ref.~\cite{Wang:1900ta} and 
too large in comparison to Ref.~\cite{Miller:1997ya}; in particular the 
input on $g_\omega$, $\kappa_\omega$~\cite{Belushkin:2006qa} has been questioned.
This criticism is unwarranted for the following reasons.  
The central value for $\kappa^{u,d}$ is due to pion-loop contributions
at a scale $M_\rho$, where changing the scale by a factor of 2 leads to a shift
of $0.008$ only.  Completely scale-independent is the next-to-leading order correction in $\kappa^{u,d}$
(which is $\Order(M_\pi)$, hence non-analytic in the quark masses), which can be read 
off as the difference between the full and the dashed curve in the magnetic form factor
in Fig.~\ref{fig:ffresults}, and which contributes roughly 40\% of the central value.
The potentially controversial vector meson contributions only
lead to the uncertainty range of $\pm 0.020$.  
Furthermore, the large anomalous $\omega N$ coupling found in 
Ref.~\cite{Belushkin:2006qa} leads to the \emph{lower} edge of the band in $G_M^{u,d}$,
Fig.~\ref{fig:ffresults}, hence to a sizeable cancellation with the pion-loop terms;
reducing these couplings makes the total result larger, not smaller.
Quite to the contrary, the quark model results for $G_M^{u,d}$~\cite{Miller:1997ya} 
are too small at low $t$ by wrongly enforcing $G_M^{u,d}(t=0)$ to vanish.

\begin{table}
\caption{Comparison of selected experimental measurements of strange form factors
from SAMPLE~\cite{Spayde:2003nr}, A4~\cite{Maas:2004dh}, and
HAPPEX~\cite{Aniol:2005zg} to the results of Ref.~\cite{Kubis:2006cy}
for the isospin-violating form factors.}
\label{tab:results}
\begin{center}
\renewcommand{\arraystretch}{1.3}
\begin{tabular}{cccc}
\hline\noalign{\smallskip}
experiment & electric/magnetic & $G^s$ & $G^{u,d}$ \\
\noalign{\smallskip}\hline\noalign{\smallskip}
SAMPLE & $G_M$            & $0.37 \pm 0.20 \pm 0.26 \pm 0.07$     & $0.02 \ldots 0.05$ \\ 
A4     & $G_E+0.106\,G_M$ & $0.071 \pm 0.036$                     & $0.004 \ldots 0.010$ \\ 
HAPPEX & $G_E+0.080\,G_M$ & $0.030 \pm 0.025 \pm 0.006 \pm 0.012$ & $0.004 \ldots 0.009$ \\
\noalign{\smallskip}\hline
\end{tabular} 
\renewcommand{\arraystretch}{1.0} 
\end{center} \vspace*{-2mm}
\end{table} 
\begin{sloppypar}
Table~\ref{tab:results} compares the specific linear combinations of
$G_E^{u,d}$ and $G_M^{u,d}$ at $t \approx -0.1\,{\rm GeV}^2$ with the 
experimentally extracted values for strangeness form factors.  
We find that the charge-symmetry-breaking shifts are still smaller 
than other experimental uncertainties, but should be kept in mind
for precision determinations of strange matrix elements.  
As another illustration, in the latest combined analysis of forward and backward asymmetries  
at A4~\cite{Baunack:2009gy}, the following values for the strange form factors
were extracted at $t=-0.22\,{\rm GeV}^2$:
\begin{equation}
G_E^s = 0.050 \pm 0.038 \pm 0.019  ~, \quad
G_M^s = -0.14 \pm 0.11 \pm  0.11   ~.
\end{equation}
Isospin breaking shifts the central values according to $0.050 \to 0.045$ for $G_E^s$ and
$-0.14 \to -0.18$ for $G_M^s$, with negligible additional errors due to the theoretical uncertainties
of $\pm 0.002$ and $\pm 0.01$, respectively.  Again, these shifts are within the given error bars,
but already of a comparable size.
\end{sloppypar}

\section{Isospin mixing in Helium-4}

Parity-violating electron scattering on ${}^4$He gives clean
access to the strange \emph{electric} form factor $G_E^s$, as the
$J^\pi=0^+$ target does not allow for magnetic or axial vector transitions.
However, in addition to effects of charge-symmetry breaking in the electric form factor
as discussed in the previous section,
an $I=1$ admixture in the ${}^4$He wave function yields a contribution
to the measured asymmetry $A_{PV}$~\cite{Viviani:2007gi}, \vspace{-1mm}
\begin{equation}
A_{PV} = - \frac{G_\mu\, t}{4\pi\alpha\sqrt{2}} 
 \Bigl\{ 4\sin^2\theta_W + \Gamma \Bigr\} ~, \quad
\Gamma = - 2\, \frac{F^{(1)}}{F^{(0)}}
- \frac{2G^{\not\,1}_E - G^s_E}{(G^p_E+G^n_E)/2} ~,
\end{equation}
where $F^{(0/1)}$ are the nuclear form factors corresponding to isoscalar/isovector
charge operators, and $G^{\not\,1}_E = ( G_{E}^{u,p} - G_{E}^{d,n} - G_{E}^{d,p} + G_{E}^{u,n})/4 $ 
is a different isospin-breaking linear combination of single-nucleon form factors.
The measured asymmetry 
$A_{PV} = \bigl[ +6.40 \pm 0.23_{\rm stat} \pm 0.12_{\rm syst} \bigr] \times 10^{-6}$
at $t=-0.077$~GeV$^2$ \cite{Acha:2006my} leads to $\Gamma = 0.010 \pm 0.038$.
Single-nucleon isospin violation contributes $0.008\pm0.003$ to $\Gamma$, 
while isospin mixing in the ${}^4$He wave function amounts to $\approx 0.003$, 
leaving a mere strangeness contribution of $G_E^s = -0.001 \pm 0.016$.  

\section{Conclusions}

We have performed an analysis of the charge-symmetry-breaking nucleon form factors,
using a combination of chiral perturbation theory and resonance saturation
that relies as far as possible on the symmetries of the Standard Model and experimental data.
Although we predict the isospin-breaking magnetic moment $G_M^{u,d}(0)$ to be different from
zero (in contrast to certain model predictions), both electric and magnetic form factors
are small, and their momentum-transfer dependence moderate.
The contributions of isospin violation to parity-violating asymmetries are as yet smaller than 
some of the experimental uncertainties in extracting strange form factors, but
clearly, charge-symmetry-breaking effects  become an essential ingredient for
precision extractions of strangeness matrix elements.

\begin{acknowledgements}
I am grateful to my coworkers Randy Lewis, Michele Viviani, and Rocco Schiavilla 
for a very fruitful collaboration, 
and further to Randy for useful comments on these proceedings.
I thank the organizers of PAVI09 for the invitation to this most enjoyable workshop. 
Partial financial support by the project ``Study of Strongly Interacting Matter'' 
(HadronPhysics2, grant 227431) under the 7th Framework Programme of the EU,
by DFG (SFB/TR 16, ``Subnuclear Structure of Matter''), and
by the Helmholtz Association providing funds to the virtual 
institute ``Spin and strong QCD'' (VH-VI-231) is gratefully acknowledged.
\end{acknowledgements}

\end{document}